%% file: main.tex
\documentclass{article}

\usepackage{microtype}
\usepackage{graphicx}
\usepackage{subcaption}
\usepackage{booktabs} 
\usepackage{tabularx}
\usepackage{hyperref}
\usepackage{multirow}
\usepackage{xcolor}
\usepackage{colortbl}
 
\usepackage{soul, xcolor}

\usepackage[accepted]{icml2026}

\usepackage{amsmath}
\usepackage{amssymb}
\usepackage{mathtools}
\usepackage{amsthm}

\usepackage[capitalize,noabbrev]{cleveref}

\theoremstyle{plain}

\theoremstyle{definition}

\theoremstyle{remark}

\usepackage[textsize=tiny]{todonotes}

\icmltitlerunning{A Geometric Perspective on Composable Emotion Steering in Text-to-Speech Models}

\begin{document}
 
\twocolumn[
  \icmltitle{A Geometric Perspective on Composable Emotion Steering in Text-to-Speech Models}
 
    \icmlsetsymbol{equal}{*}

  \begin{icmlauthorlist}
    \icmlauthor{Siyi Wang}{aff1}
    \icmlauthor{James Bailey}{aff2}
    \icmlauthor{Ting Dang}{aff1}
  \end{icmlauthorlist}

  \icmlaffiliation{aff1}{The University of Melbourne, Australia}

  \icmlaffiliation{aff2}{Monash University, Australia}

  \icmlcorrespondingauthor{Siyi Wang}{siyi.wang.4@student.unimelb.edu.au}

  \icmlkeywords{text-to-speech, activation steering, emotion control, hybrid TTS, local intrinsic dimensionality}

  \vskip 0.3in
]
 
\printAffiliationsAndNotice{}
 
\begin{abstract}
While prior work has explored emotion control in hybrid text-to-speech systems, the geometric properties of these modules, and their implications for steerability, remain poorly understood. We present the first comparative study of speech language model (SLM) and conditional flow-matching (CFM) modules as activation steering sites for mixed-emotion speech synthesis. We first characterize emotion representations using linear probing and local intrinsic dimensionality (LID), 
and then evaluate single-site and joint steering for mixed-emotion synthesis. Our results show that SLM offers a clean, low-dimensional emotion-specific subspace with strong speaker--emotion disentanglement, while CFM exhibitspoor cross-speaker generalization due to speaker--emotion entanglement.Joint steering increases emotion intensity but degrades proportional control and speech quality on in-distribution data. These findings provide practical guidance for multi-site activation steering in hybrid TTS systems and highlight the importance of representation geometry in controllable speech generation.

\end{abstract}

\input{sections/intro_new}
\input{sections/Method}

 \input{sections/experiment}
\input{sections/result-geometry}
\input{sections/results-steering}
\input{sections/discussion}
\input{sections/conclusion}

 
\bibliography{reference}
\bibliographystyle{icml2026}

\end{document}

%% file: sections/intro_new.tex
\section{Introduction}
\label{sec:intro}
Generating emotionally controllable speech is essential for applications such as conversational agents, audiobook narration, and assistive communication. Human emotional expression is nuanced, often involving mixed affective cues where multiple emotions coexist within a single utterance~\citep{zhou2022speech, cowen2017self}, a complexity that current systems generally fail to control effectively.
Existing emotion control methods operate through the model's external interface: label-based approaches can enable explicit emotion conditioning but require costly annotated data and retraining~\citep{cho2025emosphere++, gao2025emo}, while prompt-based methods can describe target emotions but lack precise quantitative control over emotion proportions~\citep{guo2023prompttts, yang2025emovoice}.

Activation steering bypasses these limitations by directly injecting learned direction vectors into intermediate activations at inference time, without retraining~\citep{zou2023representation, turner2023steering}. This paradigm has shown success in LLMs and text-to-image diffusion~\citep{rodriguez2025controlling, rimsky2024steering}. State-of-the-art TTS systems increasingly adopt hybrid architectures combining a speech language model (SLM) with a conditional flow-matching (CFM) decoder~\citep{du2024cosyvoice, anastassiou2024seed, zhou2026indextts2}, where the SLM governs high-level prosodic structure and the CFM renders fine-grained acoustics, each a potential site for steering emotional expression. \citet{wang2026cocoemo} demonstrate composable mixed-emotion steering via the SLM, while \citet{xie2025emosteer} achieve continuous single-emotion intensity control via CFM. However, no prior work has systematically compared the representation geometry at these two steering sites, examined how geometric properties relate to steering effectiveness, or investigated whether jointly steering both modules yields complementary or interfering effects.

We present the first comparative study of SLM and CFM as activation steering sites for mixed-emotion synthesis. Through linear probing and local intrinsic dimensionality (LID) analysis of both modules' representation geometry, combined with single-site and joint steering experiments on four datasets, our study reveals three key findings: (i)~the SLM encodes emotions in geometrically distinct, low-dimensional subspaces with strong cross-speaker generalization, suggesting favorable conditions for emotion-specific intervention; in contrast, the CFM entangles speaker and emotion representations, making clean emotion-only intervention difficult; (ii)~SLM steering achieves superior proportional control of mixed emotions, while CFM steering produces stronger overall intensity at the cost of speaker fidelity; (iii)~joint steering across both sites amplifies intensity but degrades proportional control in-distribution, due to two independent perturbations interfering rather than complementing each other. These findings offer practical guidelines for multi-site steering and shed light on the latent geometry of hybrid TTS, informing future work on representation control and interpretability in speech generation.

%% file: sections/Method.tex
\vspace{-5pt}
\section{Method}
\label{sec:method}

Modern hybrid TTS systems generate speech in two stages~\cite{du2024cosyvoice}. The speech language model (SLM) autoregressively generates discrete speech tokens $\mathbf{z} = f_{\text{SLM}}(\mathbf{x}, \mathbf{c}_{\text{ref}})$ from text $\mathbf{x}$ and reference audio $\mathbf{c}_{\text{ref}}$, encoding high-level prosodic and semantic structure. The conditional flow-matching (CFM) module then transforms these tokens into a mel-spectrogram $\mathbf{m} = f_{\text{CFM}}(\mathbf{z}, \mathbf{c}_{\text{ref}}, \mathbf{v})$, rendering fine-grained acoustic details. 

\vspace{-5pt}
\subsection{Geometry Analysis}
\label{sec:geometry}
The geometry analysis aims to characterize how emotions are organized in the representation spaces of SLM and CFM, and in particular whether they form structures that support compositional control. To achieve reliable mixed-emotion steering, individual emotion directions should ideally be composable, so that their weighted combinations produce meaningful mixed directions~\citep{wang2026cocoemo}.

\vspace{-8pt}
\paragraph{Linear Discriminability}
\label{sec:linear_method}
We use linear probing to analyze the linear separability of emotion representations in the activation space~\cite{alain2016understanding}. Concretely, we train a linear classifier at each layer of the SLM and CFM, and compare their classification performance across layers. Higher classification performance indicates more separable emotion representations, and possibly more reliable steering vectors for compositional steering
~\citep{wang2026cocoemo}.

\vspace{-8pt}
\paragraph{Local Intrinsic Dimensionality}
\label{sec:lid_method}

To further reveal the geometric structure of the representation manifold, Local intrinsic dimensionality (LID) is used~\citep{amsaleg2015estimating}. For each sample’s activation representation, we compute its $K$ nearest neighbors (in Euclidean distance) in the activation space. Let $r_1, r_2, \cdots, r_K$ denote the distances to these neighbors sorted in ascending order. We then estimate the LID by modeling the growth rate of the neighborhood radius using the Levina--Bickel maximum likelihood estimator~\cite{levina2004maximum}, which captures how quickly the local volume expands around each point. Higher LID indicates a more complex and less constrained local geometry, suggesting that emotion information is distributed across a higher-dimensional space rather than in a compact subspace.

We compute LID for each emotion as well as over all speech samples. The \emph{per-emotion} setting captures the geometry of each emotion-specific subspace, while the all-samples setting captures the overall geometry of the full emotion space (referred to as the \emph{pooled} setting). 
We define $\Delta\text{LID} = \text{LID}_{\text{pooled}} - \overline{\text{LID}}_{\text{per-emo}}$ as the difference between pooled and average of per-emotion LID. When $\Delta\text{LID} > 0$, pooling emotions increases the estimated manifold dimensionality, indicating that different emotions contribute additional independent directions of variation beyond those captured within individual emotion subspaces. Conversely, when $\Delta\text{LID} < 0$, pooling does not increase dimensionality, suggesting that emotion categories largely lie on a shared manifold. A positive $\Delta\text{LID}$ is favorable for mixed-emotion steering, as it indicates emotion-specific directions in the representation space that can potentially be composed.

\vspace{-8pt}
\subsection{Activation Steering}
\label{sec:steering_method}
Building on these geometric analyses, we next turn to how emotion directions are extracted from activations and used for steering. For each layer $l$ at either SLM or CFM, we first extract the activation difference for each emotion $e$ as:

\vspace{-10pt}
\begin{equation}
\small
\mathbf{u}_e^{(l)} = \frac{1}{N_e}\sum_{j=1}^{N_e} \mathbf{h}_{e,j}^{(l)} - \frac{1}{N_0}\sum_{i=1}^{N_0} \mathbf{h}_{0,i}^{(l)}, 
\vspace{-10pt}
\end{equation}
where $\mathbf{h}_{e,j}^{(l)}$ and $\mathbf{h}_{0,i}^{(l)}$ denote activations from emotion-$e$ and neutral samples, respectively. 
For the SLM, steering vector $\mathbf{v}_e^{(l)} = \mathbf{u}_e^{(l)}$ is extracted from attention output activations at the last-token position of complete utterances
~\citep{wang2026cocoemo}. For the CFM, $\mathbf{u}_e^{(l)}$ is extracted from residual stream activations, $L_2$-normalized, masked to the top-$k$ emotion-relevant frames identified via an emotion classifier, and aggregated to get $\mathbf{v}_e^{(l)}$
~\citep{xie2025emosteer}. For mixed-emotion synthesis, single-emotion vectors are composed via weighted summation: $\mathbf{v}_{\text{mix}}^{(l)} = \sum_e p_e \mathbf{v}_e^{(l)}$, where $p_e$ denotes the proportion summing to 1.

At inference, steering is applied by modifying the activation 
at layer $l$:
$\tilde{\mathbf{h}}^{(l)} = f_r(\mathbf{h}^{(l)} + \alpha \cdot \mathbf{v}_{\text{mix}}^{(l)})$,
where $\alpha$ controls steering strength and $f_r$ renormalizes the modified activation to preserve the original scale~\citep{turner2023steering}.

%% file: sections/experiment.tex
\vspace{-5pt}
\section{Experimental Setup}
\label{sec:setup}

\begin{figure*}[t]
  \centering

  \begin{subfigure}[t]{0.33\textwidth}
    \centering
    \includegraphics[width=\linewidth]{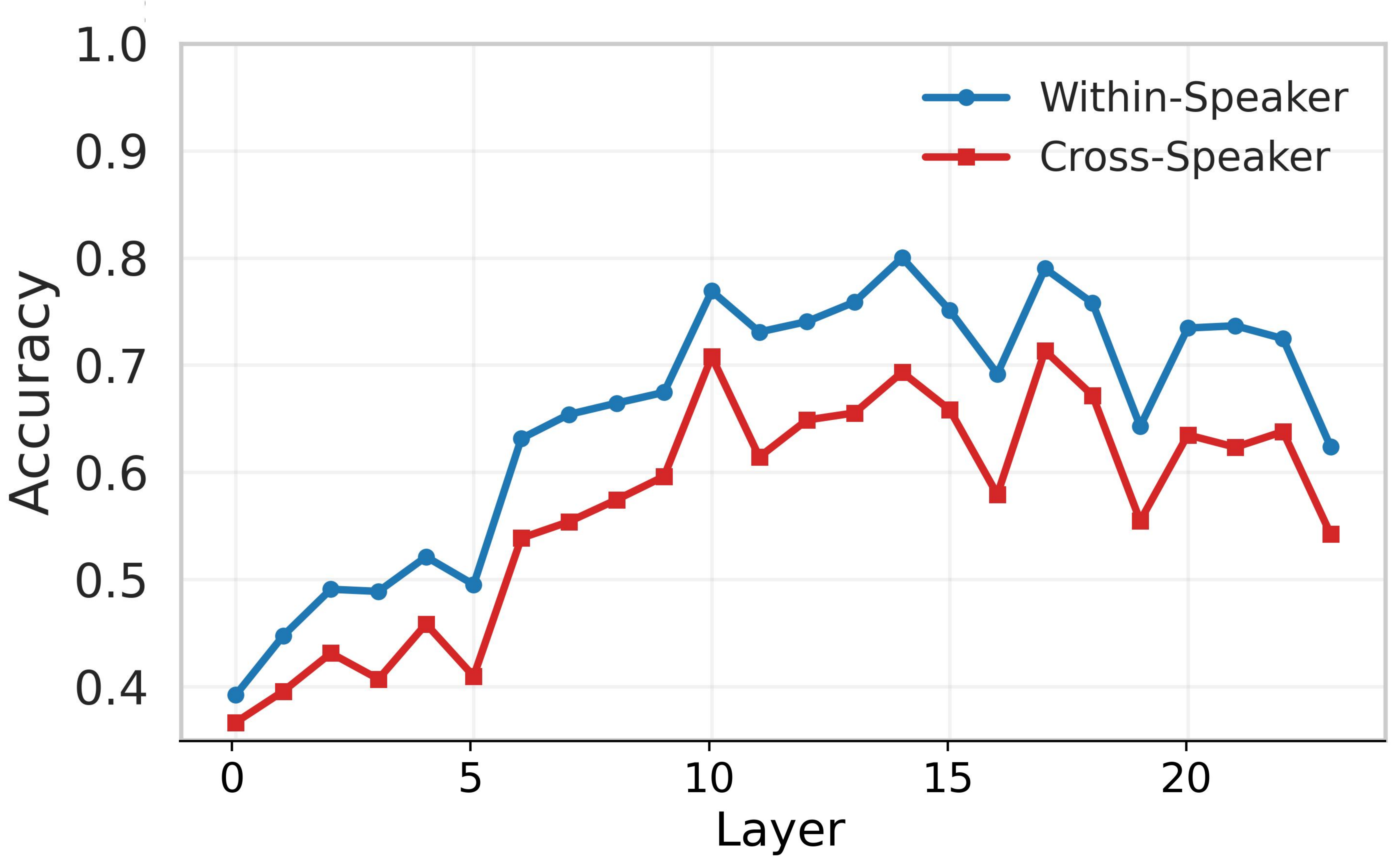}
    \caption{Emotion discriminability in SLM}
    \vspace{-8pt}
    \label{fig:SLM_linear_discriminability}
  \end{subfigure}%
  \hfill
  \begin{subfigure}[t]{0.33\textwidth}
    \centering
    \includegraphics[width=\linewidth]{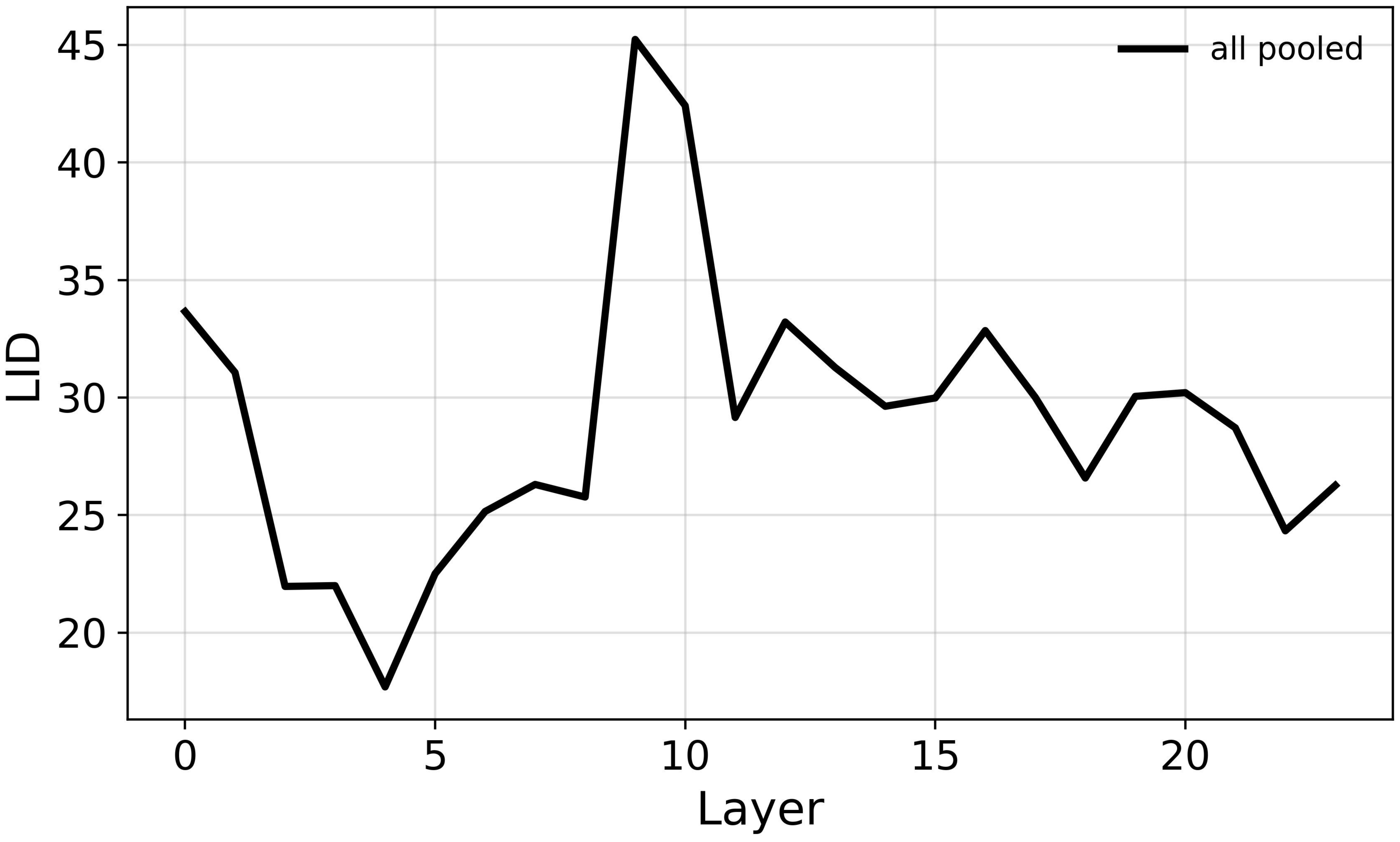}
        \caption{Pooled LID in SLM}
    \vspace{-8pt}
    \label{fig:SLM_pool_LID}
  \end{subfigure}%
  \hfill
  \begin{subfigure}[t]{0.33\textwidth}
    \centering
    \includegraphics[width=\linewidth]{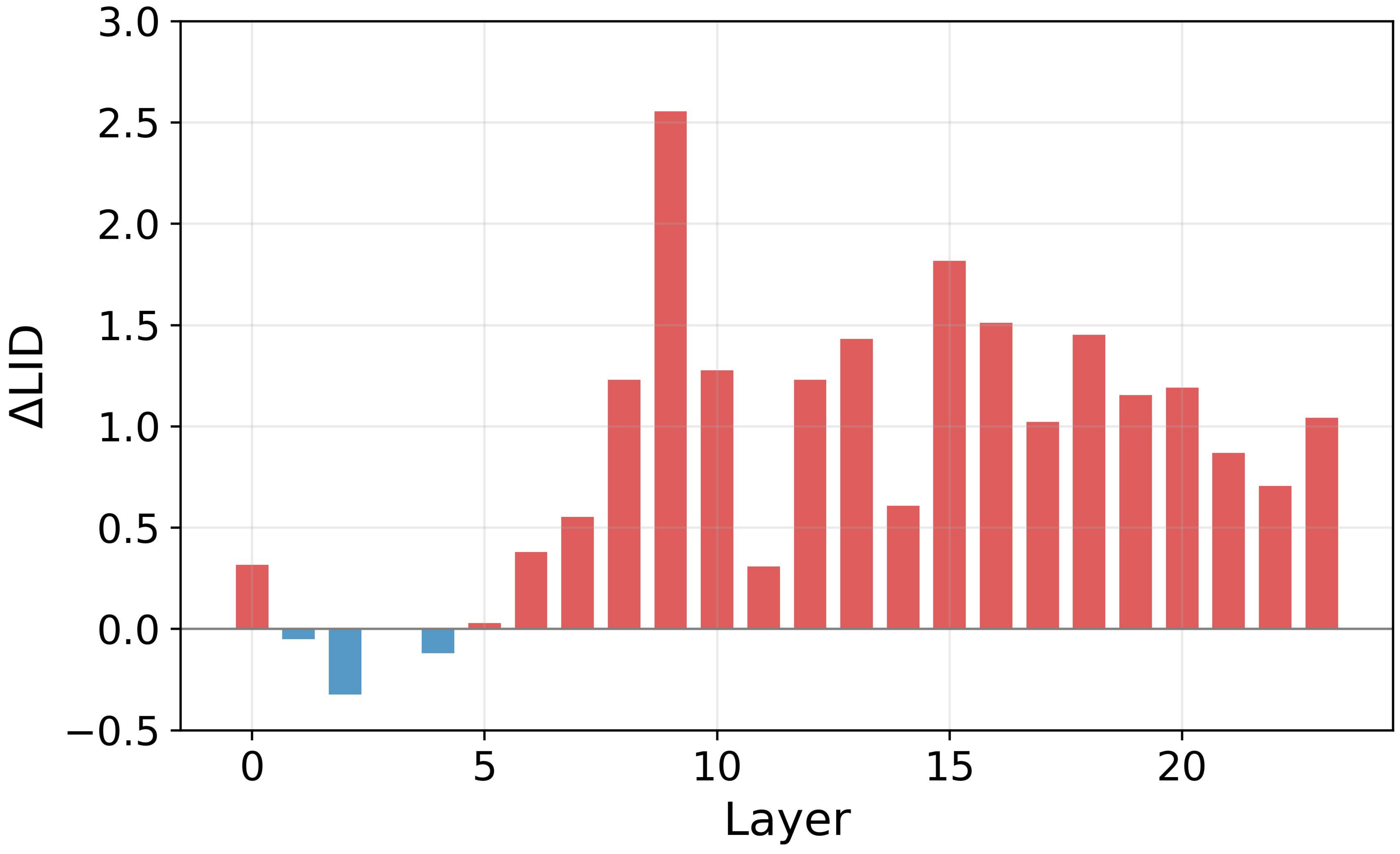}
    \caption{$\Delta$LID in SLM}
    \vspace{-8pt}
    \label{fig:SLM_delta_LID}
  \end{subfigure}

  \vspace{1em}  

  \begin{subfigure}[t]{0.33\textwidth}
    \centering
    \includegraphics[width=\linewidth]{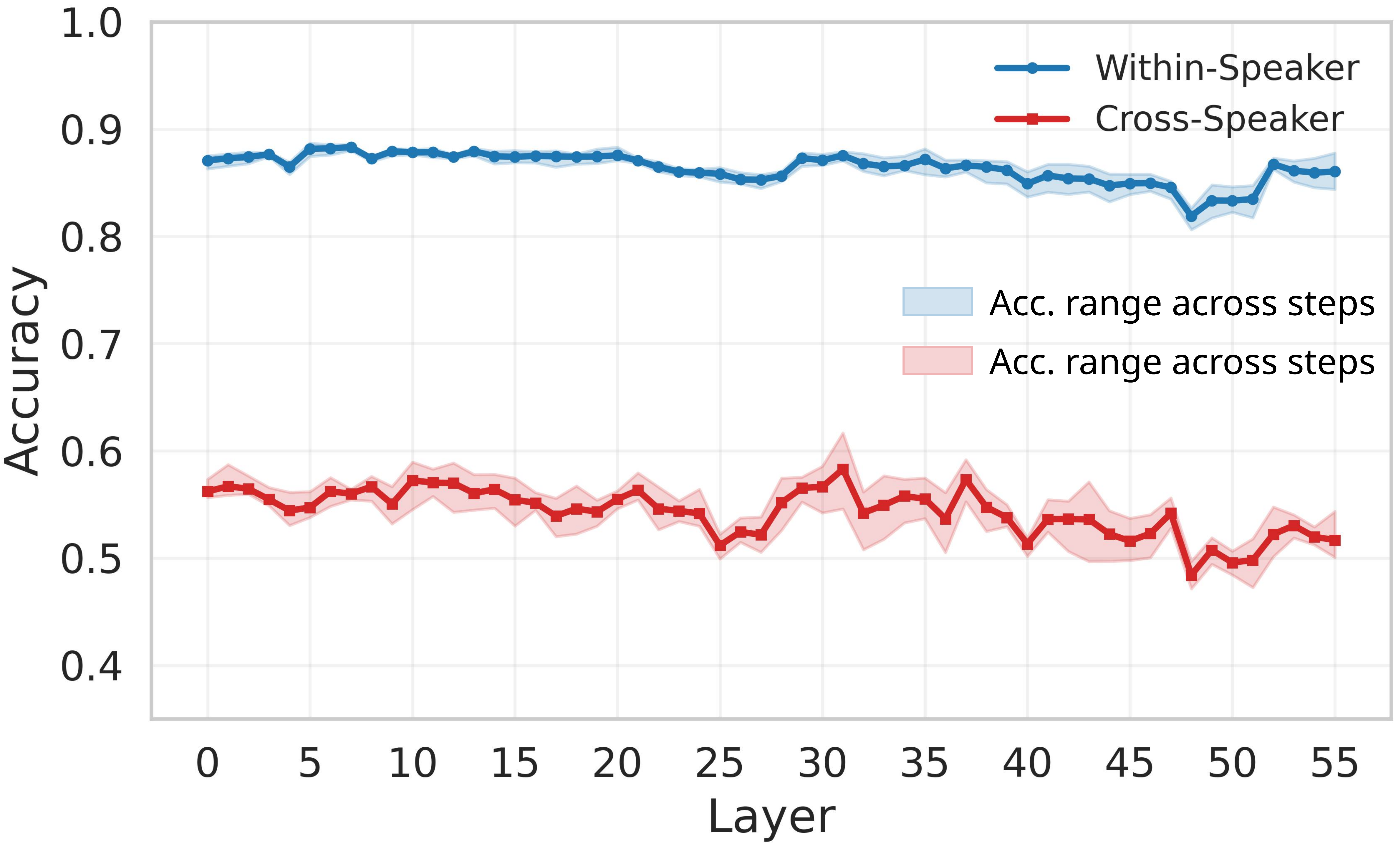}
    \caption{Emotion discriminability in CFM }
    \vspace{-3pt}
    \label{fig:CFM_linear_discriminability}
  \end{subfigure}%
  \hfill
  \begin{subfigure}[t]{0.32\textwidth}
    \centering
    \includegraphics[width=\linewidth]{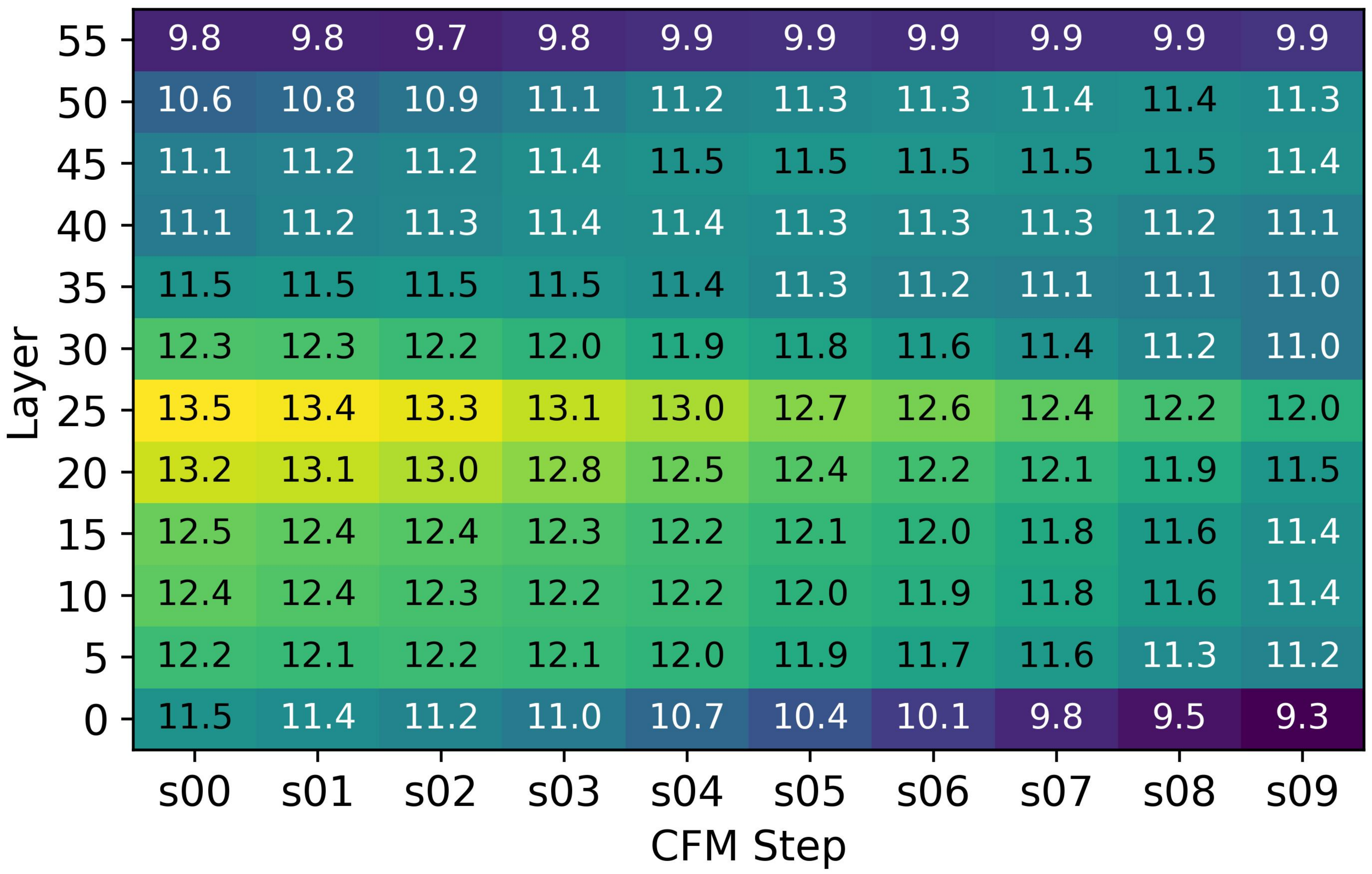}
    \caption{Pooled LID in CFM}
    \vspace{-3pt}
    \label{fig:CFM_pool_LID}
  \end{subfigure}%
  \hfill
  \begin{subfigure}[t]{0.32\textwidth}
    \centering
    \includegraphics[width=\linewidth]{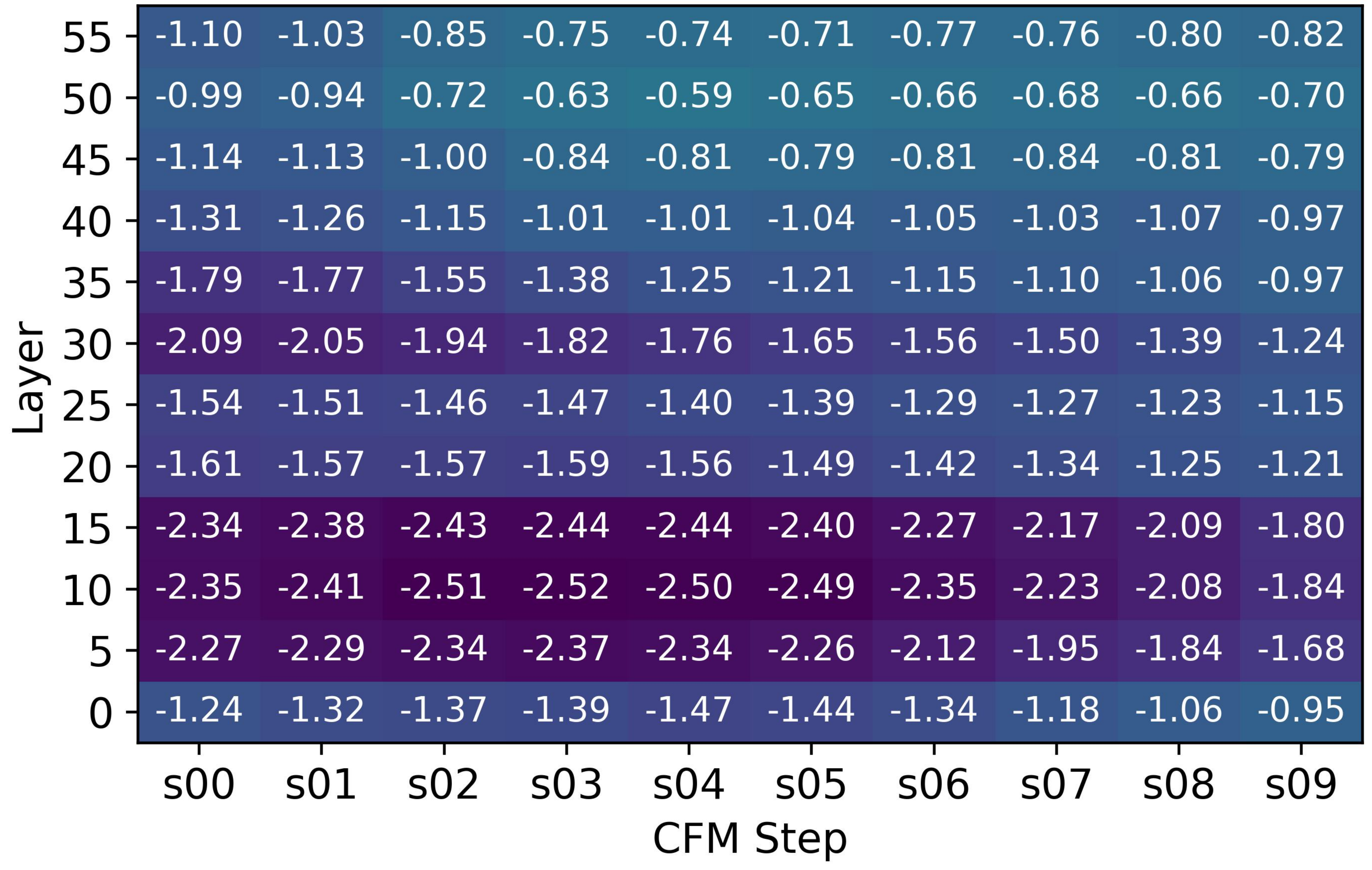}
    \caption{$\Delta$LID in CFM}
    \vspace{-3pt}
    \label{fig:CFM_delta_LID}
  \end{subfigure}
  \caption{
  Geometry analysis of SLM (a–c) and CFM (d–f). (a,d) Per-layer emotion discriminability (linear-probe accuracy); blue = within-speaker, red = cross-speaker; shading in (d) shows the accuracy range across denoising steps. (b,e) Pooled LID across layers (CFM shown per denoising step). (c,f) $\Delta\text{LID} = \text{LID}_{\text{pooled}} - \overline{\text{LID}}_{\text{per-emo}}$ across layers.}
  \vspace{-15pt}
  \label{fig:all_six}
\end{figure*}

\textbf{Model.} We use CosyVoice2~\citep{du2024cosyvoice} as our backbone. The SLM is a 24-layer Qwen2.5-based~\cite{qwen2025qwen25technicalreport} transformer and the CFM is a 56-layer DiT with 10 denoising steps. For geometry analysis, we extract activations from all SLM layers and CFM layers across 10 steps. 
For steering, based on the findings in Section~\ref{sec:geometry_comparison}, we apply steering vectors at SLM layers 14 and 17~\citep{wang2026cocoemo}. For CFM steering, since emotion discriminability is uniformly distributed across layers (Section~\ref{sec:geometry_comparison}), we follow \citet{xie2025emosteer} and apply steering vectors at every 5th layer (12 layers in total) across all 10 denoising steps.

\textbf{Datasets.} We use ESD~\citep{zhou2021seen}, CREMA-D~\citep{cao2014crema}, and RAVDESS~\citep{livingstone2018ryerson} across five emotions (angry, happy, neutral, sad, surprise). For linear probing, we reserve 30\% of speakers for cross-speaker evaluation (4,530 utterances). From the remaining speakers, 11,311 utterances are used for probe training and 4,850 utterances for within-speaker evaluation. For LID, we sample 4,000 utterances for both per-emotion and pooled estimates, with $k{=}50$ neighbors, averaged over 10 resampling trials. Steering vectors are extracted from 50\% of speakers and evaluated on CREMA-D (in-distribution) and IEMOCAP~\citep{busso2008iemocap} (out-of-distribution), with mixed-emotion ground truths derived from multi-rater annotation disagreement~\citep{wang2026cocoemo}.

\textbf{Evaluation metrics.} We evaluate along two axes. For emotion control we use: \textbf{E-SIM}, cosine similarity between Emotion2Vec embeddings~\citep{ma2024emotion2vec} of synthesized and ground-truth speech; \textbf{TEP}, mean probability assigned to target emotions by the Emotion2Vec classifier; \textbf{$\rho$}, Spearman correlation between the ranking of emotion probability increases and the ground-truth emotion ranking; and \textbf{H-Rt}, fraction of samples where the ground-truth dominant emotion shows the largest probability increase. For speech quality we report: \textbf{S-SIM}, cosine similarity between WavLM speaker embeddings~\citep{chen2022wavlm} of synthesized and reference speech; and \textbf{WER}, word error rate via Whisper-Large-v3~\citep{radford2023robust}.

%% file: sections/result-geometry.tex
\vspace{-5pt}
\section{Results}

\subsection{Geometry Comparison}
\label{sec:geometry_comparison}

\textbf{Linear discriminability.}
In the SLM, within-speaker accuracy reaches 0.80 and cross-speaker accuracy 0.71, yielding a small mean gap of 0.08 (Figure~\ref{fig:SLM_linear_discriminability}). Discriminability peaks in the mid-to-late layers (10--17), indicating that emotion information is concentrated in a localized, speaker-invariant subspace. In contrast, the CFM achieves similarly high within-speaker accuracy (0.89) but much lower cross-speaker accuracy (0.62), resulting in a larger mean gap of 0.32 (Figure~\ref{fig:CFM_linear_discriminability}). Moreover, discriminability is broadly uniform across layers and denoising steps, with no clear peak. These results suggest that SLM representations are more separable and generalizable across speakers, while CFM representations are more entangled with speaker identity and diffusely distributed, making the SLM a more suitable site for extracting robust emotion steering directions.

\textbf{LID trend.}
In the SLM (Figure~\ref{fig:SLM_pool_LID} black line), pooled LID exhibits a compression–expansion pattern, first decreasing, then increasing, and finally stabilizing in later layers, consistent with geometric dynamics observed in transformer representations~\citep{valeriani2023geometry}. In contrast, the CFM (Figure~\ref{fig:CFM_pool_LID}) consistently exhibits an increase followed by a decrease in LID across layers at every denoising step, suggesting that intermediate layers construct richer representations with more complex local geometry before compressing them toward the prediction target. Furthermore, LID progressively decreases across denoising steps, indicating that the representation manifold becomes increasingly structured and lower-dimensional as CFM iteratively refines towards final output~\citep{lipman2022flow}.

\textbf{Emotion subspace structure ($\Delta$LID).}
$\Delta$LID reveals a fundamental contrast between the two modules (Figure~\ref{fig:SLM_delta_LID},\ref{fig:CFM_delta_LID}). In the SLM, $\Delta$LID is near zero in early layers (0--5), then becomes consistently positive from layer 6 onward (mean: $+0.84$), indicating that combining emotion categories increases manifold dimensionality, and that emotions occupy distinct directions and contribute additional geometric structure beyond that of individual emotion manifolds. 
In the CFM, $\Delta$LID is negative across all 56 layers and 10 steps (mean: $-1.48$), indicating that pooling emotions does not increase dimensionality and that emotion categories largely reside on a shared acoustic manifold.
This contrast suggests that SLM contains more distinct emotion subspaces and is more favorable for compositional steering.
The overall comparison is summarized in Table~\ref{tab:geometry}.

\begin{table}[t]
\caption{Geometric comparison of SLM and CFM as steering sites on CosyVoice2. 
Probe accuracies are best-layer values; within–cross gap is averaged across all layers and steps.}
\vspace{-5pt}
\label{tab:geometry}
\centering
\small
\setlength{\tabcolsep}{4pt}
\begin{tabular}{lcc}
\toprule
\textbf{Property} & \textbf{SLM} & \textbf{CFM} \\
\midrule
Hidden Dim. & 896 & 256\\
Probe acc. (within / cross) & 0.80 / 0.71 & 0.89 / 0.62 \\
Mean within--cross gap & 0.08 & 0.32 \\
Manifold dim. (LID) & ${\sim}$28 & ${\sim}$13 \\
$\Delta$LID & Positive (+0.84) & Negative ($-$1.48) \\
Discriminability peak & Mid-to-late & Uniform \\
\bottomrule
\end{tabular}
\end{table}
 
\vspace{-10pt}

%% file: sections/results-steering.tex
\subsection{Steering Comparisons}
\label{sec:steering_comparisons}

\begin{table}[t]
\caption{Steering results for mixed-emotion speech synthesis on CosyVoice2. Best in \textbf{bold}, second \underline{underlined}.}
\vspace{-5pt}
\label{tab:steering}
\centering
\small
\setlength{\tabcolsep}{2pt}
\begin{tabular}{ll cccccc}
\toprule
\textbf{Data} & \textbf{Config} & \textbf{E-SIM}$\uparrow$ & \textbf{TEP}$\uparrow$ & $\boldsymbol{\rho}\uparrow$ & \textbf{H-Rt}$\uparrow$ & \textbf{S-SIM}$\uparrow$ & \textbf{WER}$\downarrow$ \\
\midrule
\multirow{7}{*}{\rotatebox{90}{CREMA-D}}
 & No-steer          & .743 & .065 & --   & --   & \underline{.871} & 1.07 \\
 & CFM \tiny{$\alpha\!=\!1.0$}  & .767 & .097 & .098 & .691 & .858 & \textbf{0.76} \\
 & CFM \tiny{$\alpha\!=\!2.0$}   & \underline{.786} & \underline{.160} & \underline{.193} & \underline{.717} & .807 & 0.79 \\
 & SLM \tiny{$\alpha\!=\!3.0$}   & .762 & .100 & .166 & .709 & \textbf{.872} & 1.01 \\
 & SLM \tiny{$\alpha\!=\!5.0$}   & .779 & .149 & \textbf{.209} & \textbf{.724} & .870 & \underline{0.78} \\
\cmidrule{2-8}
 & Joint \tiny{$\alpha\!=\!1.0$} & .767 & .131 & .112 & .695 & .859 & 1.02 \\
 & Joint \tiny{$\alpha\!=\!2.0$} & \textbf{.787} & \textbf{.163} & .176 & .711 & .808 & 1.06 \\
\midrule
\multirow{7}{*}{\rotatebox{90}{IEMOCAP}}
 & No-steer          & .903 & .197 & --   & --   & .888 & 6.70 \\
 & CFM \tiny{$\alpha\!=\!1.0$}   & .910 & .218 & .138 & .729 & .885 & 6.08 \\
 & CFM \tiny{$\alpha\!=\!2.0$}   & .909 & \underline{.272} & .117 & .721 & .844 & 6.15 \\
 & SLM \tiny{$\alpha\!=\!3.0$}   & .911 & .228 & .186 & .744 & \textbf{.891} & \textbf{5.86} \\
 & SLM \tiny{$\alpha\!=\!5.0$}  & \textbf{.915} & .253 & \textbf{.215} & \textbf{.755} & \underline{.890} & 6.27 \\
\cmidrule{2-8}
 & Joint \tiny{$\alpha\!=\!1.0$} & \underline{.912} & .237 & \underline{.193} & \underline{.746} & .884 & \underline{6.05} \\
 & Joint \tiny{$\alpha\!=\!2.0$} & .911 & \textbf{.274} & .170 & .737 & .845 & 6.29 \\
\bottomrule
\end{tabular}
\vspace{-10pt}
\end{table}

Table~\ref{tab:steering} compares steering applied at the SLM only, CFM only, and both modules jointly. Our SLM-only and CFM-only conditions instantiate the steering approaches of~\cite{wang2026cocoemo} and~\cite{xie2025emosteer} respectively. For each steering site, we vary the steering strength $\alpha$ and report the best-performing configuration under comparable preserved speech quality (S-SIM within 10\% of baseline, WER increase ${<}0.5$).

\textbf{Emotion control.}
Emotion embedding similarity (E-SIM) and target emotion intensity (TEP) improve over the baseline for both SLM and CFM steering, with comparable performance, indicating that both sites effectively align generated speech with target emotional embeddings. 
Joint steering yields the highest TEP across datasets, as combined perturbations reinforce overall emotion intensity.

For fine-grained proportional control of each emotion ($\rho$, H-Rt), SLM steering consistently outperforms CFM on both datasets (Table~\ref{tab:steering}), consistent with geometric analysis. The positive $\Delta$LID and low-dimensional emotion subspace in SLM support cleaner compositional steering, enabling the precise control over emotion mixing. However, joint steering degrades proportional control on in-distribution data, suggesting that steering both modules simultaneously complicates the control over individual emotion ratios.

\textbf{Speech quality cost.}
S-SIM degrades noticeably under CFM steering, while SLM steering preserves speaker identity. This is consistent with CosyVoice2 architecture: the SLM is not conditioned on speaker embeddings, whereas the flow-matching module is explicitly conditioned on speaker embeddings and reference speech~\citep{du2024cosyvoice}, so perturbing CFM activations directly interferes with speaker-dependent representations. This aligns with the speaker--emotion entanglement revealed by geometry analysis (\S\ref{sec:geometry_comparison}). WER remains stable for single-site methods but increases slightly under joint steering. 
Overall, \emph{SLM steering provides a better balance between controllability and preservation, making it the more suitable site for emotion steering}.

%% file: sections/discussion.tex
\noindent\textbf{Steering analysis. }

The results in \S\ref{sec:steering_comparisons} show that joint steering does not provide additive gains over single-site steering, which we attribute to three factors. (i) Distribution shift: SLM steering moves activations away from the neutral manifold before they reach the CFM module, causing a mismatch for CFM steering vectors, especially in mixed-emotion settings. (ii) Speaker entanglement: CFM steering additionally perturbs speaker-dependent acoustics due to speaker–emotion entanglement in the flow-matching space. (iii) Uncoordinated perturbation: independent interventions at both sites accumulate noise rather than compose, leading to interference that reduces proportional control despite increasing overall emotion intensity.

\noindent\textbf{Future directions.}
(i)~CFM vectors could be extracted conditioned on SLM-steered output so that the extraction distribution matches actual inference-time conditions.
(ii)~Steering vectors in the CFM could be orthogonalized against speaker directions~\citep{ravfogel2020null, bartoszcze2025representation} to mitigate speaker--emotion entanglement.
(iii)~Independent per-site $\alpha$ tuning or frame-level adaptive steering may allow the two modules to complement each other more effectively.
(iv)~Extending this analysis to architecturally distinct systems (e.g., IndexTTS2) would test generality.
(v)~ The geometry–steering relationship, currently characterized at the module level, could be analyzed per-layer and per-step to pinpoint the most steerable directions within each module.

%% file: sections/conclusion.tex
\vspace{-5pt}
\section{Conclusion}
 
We presented the first comparative study of SLM and CFM modules as activation steering sites in hybrid TTS, providing a geometry-to-application analysis of how each module encodes and controls emotion. Our findings reveal distinct roles: the SLM provides clean, speaker-invariant emotion subspaces suited for proportional mixed-emotion control, while the CFM module contributes rich acoustic detail but entangles emotion with speaker identity. Joint steering amplifies intensity but introduces interference, highlighting the need for coordinated multi-site strategies. By connecting representation geometry to steering outcomes, this work offers both an analytical framework and practical guidance for emotion control in hybrid TTS architectures.